\documentclass[aps,prb,reprint,longbibliography]{revtex4-2}

\usepackage{graphicx}
\usepackage{siunitx}
\usepackage{bbm}
\usepackage{amsmath,amsfonts,amssymb,braket,bbold,mathtools,array}
\usepackage[colorlinks,citecolor=blue,linkcolor=blue,urlcolor=blue,bookmarks=false,hypertexnames=true]{hyperref}

\newcommand{\eps}{\varepsilon}
\renewcommand{\vec}[1]{\boldsymbol{#1}}
\newcommand{\df}{\delta\varphi}
\newcommand{\ain}{a_\mathrm{in}}
\newcommand{\aout}{a_\mathrm{out}}

\begin{document}

\title{Proposal for a cavity-induced measurement of the exchange coupling in quantum dots}

\author{Florian Ginzel}
\author{Guido Burkard}
\affiliation{Department of Physics, University of Konstanz, D-78457 Konstanz, Germany}


\begin{abstract}
In spin qubit arrays the exchange coupling can be harnessed to implement two-qubit gates and to realize intermediate-range qubit connectivity along a spin bus. In this work, we propose a scheme to characterize the exchange coupling between electrons in adjacent quantum dots. We investigate theoretically the transmission of a microwave resonator coupled to a triple quantum dot (TQD) occupied by two electrons. We assume that the right quantum dot (QD) is always occupied by one electron while the second electron can tunnel between the left and center QD. If the two electrons are in adjacent dots they interact via the exchange coupling. By means of analytical calculations we show that the transmission profile of the resonator directly reveals the value of the exchange coupling strength between two electrons. From perturbation theory up to second order we conclude that the exchange can still be identified in the presence of magnetic gradients. A  valley splitting comparable to the inter-dot tunnel coupling will lead to further modifications of  the cavity transmission dips that also depend on the valley phases.
\end{abstract}

\maketitle

\section{Introduction\label{sec_intro}}

Spin qubits in quantum dots (QDs) \cite{PhysRevA.57.120} promise to be an excellent quantum information platform. The choice of silicon as host material with its abundant nuclear spin-free $^{28}$Si isotope has granted remarkably long coherence times \cite{Si28T2star,EnrichedSilicon}. 
Two-qubit gates can be performed utilizing the exchange coupling between electrons in neighboring QDs \cite{PhysRevA.57.120,PhysRevB.83.121403,Zajac439,PhysRevB.97.085421,Watson2018}. The exchange coupling can also serve as a backbone of intermediate range qubit interaction via a spin bus \cite{PhysRevLett.96.247206,PhysRevLett.98.230503,Sigillito2019,doi:10.1021/acs.nanolett.0c04771,PhysRevX.8.011045} and allows to advance beyond the original proposal of a natural spin-$1/2$ qubit \cite{DiVincenzo2000}. Multi-spin qubits in exchange-coupled quantum dots \cite{exchangeonlyqubit,ST0-Qubit,Petta2180,PhysRevLett.111.050502,PhysRevLett.111.050501,PhysRevLett.121.177701,PhysRevB.95.241303} are more complex but provide benefits in terms of stability and control. Typical ways to characterize the exchange interaction in a given device include transport experiments \cite{Lai2011,GolovachThesis}, spin-funnel measurements \cite{Petta2180,Maune2012} and other types of gate-based spectroscopy \cite{PhysRevX.8.011045,Chen2020}.

The integration of spin qubits into circuit quantum electrodynamics architectures \cite{GB_review} has further advanced the scalability of QD implementations \cite{vandersypen_scaling,Lieaar3960,Holman2020,GonzalezZalba2020}. The electric dipole coupling can establish a coherent interface between microwave resonator photons and the charge \cite{Mi156} and spin \cite{PhysRevB.86.035314,StrongCouplingPrinceton,Samkharadze1123,Landig2018} degree of freedom of a confined electron in a double quantum dot (DQD). This allows coupling between distant qubits \cite{PhysRevA.69.042302,PhysRevB.74.041307,PhysRevB.97.235409,Benito2019,Borjans2020}. Furthermore, by injecting a probe field into the resonator and monitoring the output field it is possible to read out the qubit state \cite{Peterson2010,PhysRevLett.110.046805,House2015,PhysRevB.100.245427,Crippa2019,Vandersypen_readout,PhysRevX.10.041010} and to investigate the electronic energy spectrum \cite{doi:10.1021/acs.nanolett.5b01306,PhysRevLett.122.213601}. In particular, resonators coupled to silicon QDs can successfully aid the characterization the valley Hamiltonian \cite{PhysRevB.94.195305,PhysRevLett.119.176803,Russ_2020,PhysRevB.98.161404,Hao2014,Borjans2021,Chen2021} which is hard to access otherwise.
The valley degree of freedom is a potential complication for silicon spin qubits due to the six-fold degenerate conduction band minimum of silicon \cite{RevModPhys.54.437,PhysRevB.20.734}. The QD confinement potential partially lifts the degeneracy into the additional valley pseudospin and a split-off manifold with higher energy \cite{doi:10.1063/1.1637718,PhysRevLett.88.027903,PhysRevB.81.115324,Hollmann2019}. The valley Hamiltonian strongly depends on the microscopic environment \cite{PhysRevB.80.081305,PhysRevB.84.155320,PhysRevResearch.2.043180}. The hybridization of valley, orbital and spin states \cite{PhysRevB.98.161404} may give rise to unwanted effects such as enhanced relaxation near the spin-valley hotspot \cite{PhysRevB.71.205324,PhysRevB.72.155410,PhysRevLett.110.196803,PhysRevApplied.14.054015}, lifted Pauli blockade \cite{PhysRevB.82.155312,Hao2014} or an altered probability distribution of spin measurements \cite{Buterakos2021}.

This raises the question whether microwave cavity transmission can be used to probe the exchange coupling between neighboring quantum dots.
At the surface, it appears that the answer is negative because, despite its many advantages, a coupled microwave resonator is not well-suited to measure the exchange coupling between two electrons in a DQD. This is because the exchange interaction emerges deep in the (1,1) charge sector where the electron number is fixed and charge transitions between the two QDs are extremely unlikely. The electric dipole of a DQD in this regime and hence the coupling to the cavity field are extremely low.
\begin{figure}
\begin{center}
\includegraphics[width=0.5\textwidth]{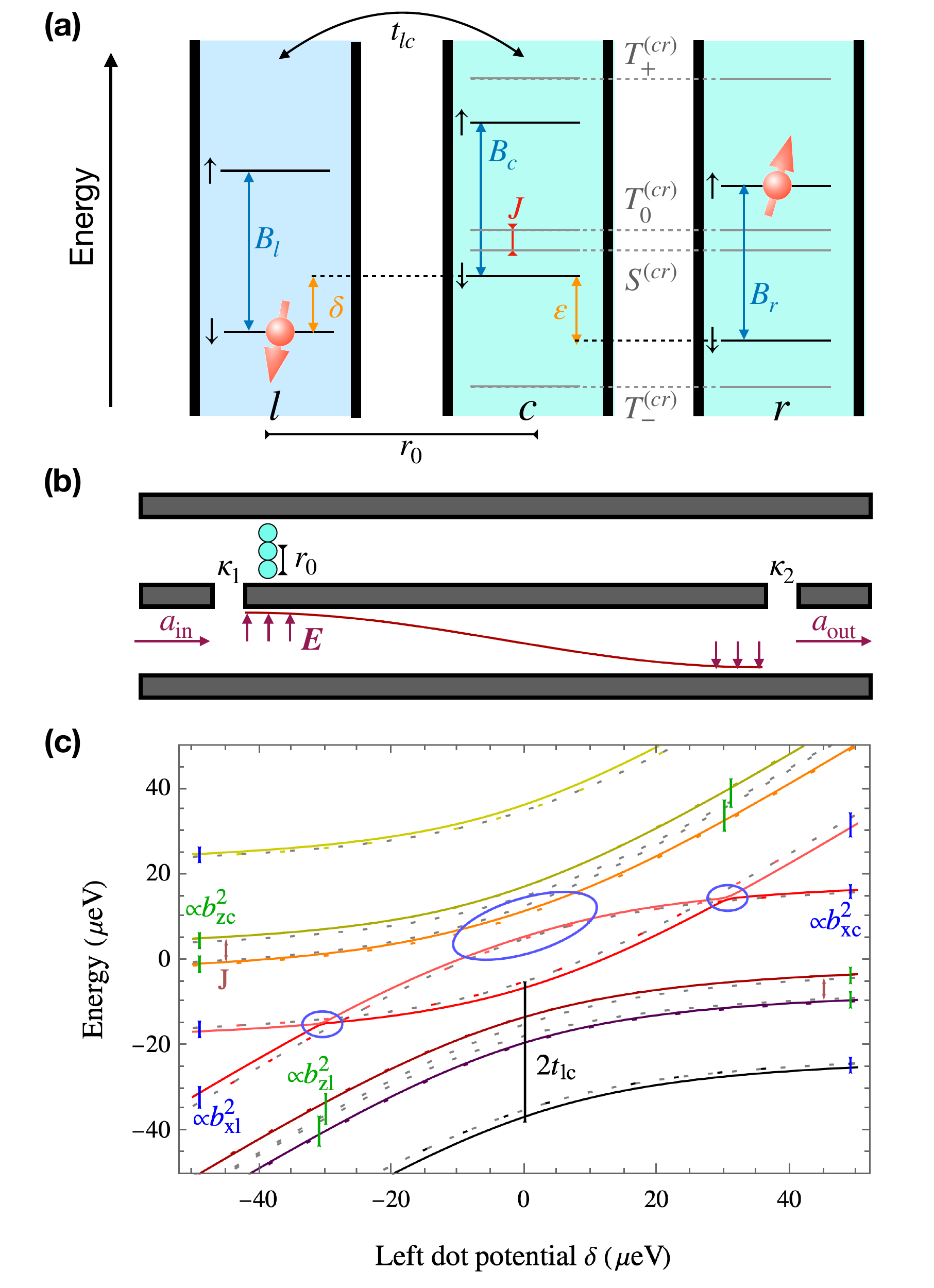}
\caption{Schematic depiction of the system.
(a) Energy levels of the TQD. The on-site potentials are assumed to be set such that the right dot ($r$) is permanently occupied by one electron while a second electron can tunnel between the left ($l$) and center ($c$) dot. In the $(0,1,1)$ charge configuration the exchange energy $J$ between singlet and triplet states emerges due to the wavefunction overlap. The single-electron physics of the TQD is characterized by Zeeman splittings $\vec{B}_l$, $\vec{B}_c$, $\vec{B}_r$, tunneling $t_{lc}$ and energy detunings $\eps$, $\delta$.
(b) Sketch of a microwave resonator with embedded TQD. The electric field couples to the dipole moment associated with the charge transition $(1,0,1)\leftrightarrow (0,1,1)$. (c) Energy levels of the TQD as a function of $\delta$ near the charge transition. Dashed gray lines are plotted with $J=\SI{5}{\micro\electronvolt}$, $t_{lc}=\SI{15}{\micro\electronvolt}$,  $B_\mathrm{ext}=\SI{20}{\micro\electronvolt}$, $\eps=\SI{100}{\micro\electronvolt}$ and without magnetic gradients. The dashed colored lines include a longitudinal gradient of $b_{zl}=2b_{zc}=\SI{4}{\micro\electronvolt}$ which hybridizes the singlet and $T_0$ states (indicated in green).
The solid curves additionally include a transverse gradient $b_{xl}=2b_{xc}=\SI{10}{\micro\electronvolt}$ which introduces spin-flip processes. Thus, the  polarized triplets $T_\pm$ are shifted in energy and  further avoided crossings are opened (indicated in blue).\label{fig_systemsketch}}
\end{center}
\end{figure}

In this paper, we show that in fact the exchange coupling can be probed by a cavity in a triple quantum dot (TQD). We study theoretically the transmission of a microwave resonator coupled to a TQD occupied by two electrons close to the $(1,0,1)\leftrightarrow (0,1,1)$ charge transition (Fig.~\ref{fig_systemsketch}). Here, $(n_l,n_r,n_c)$ denote the numbers of electrons in the left ($l$), center ($c$) and right ($r$) QD. In the $(0,1,1)$ configuration the short-ranged exchange interaction couples the two electrons and splits the spin singlet and triplet states in energy. The resonator transmission $T$ exhibits a response to the avoided level crossings of the charge transition. We show that it is possible to extract the exchange coupling $J$ from the cavity response during a sweep of the left dot potential. We further derive the effect of an inhomogeneous magnetic field on $T$ and show that the measurement scheme also works in the presence of a lifted valley degeneracy.

This article is organized as follows. In Sec.~\ref{sec_TQD_model} we introduce the model for the coupled TQD and resonator. In Sec.~\ref{sec_simplest_case} we describe the cavity-induced measurement of the exchange interaction. In Secs.~\ref{sec_grad} and \ref{sec_valley}, we discuss the effect of magnetic gradients and a lifted valley degeneracy on the transmission profile. In Sec.~\ref{sec_conclusion} we summarize our results.

\section{Cavity-coupled TQD model\label{sec_TQD_model}}

To model the 
TQD we introduce the Hamiltonian $H_\mathrm{TQD}$ which incorporates the electrostatic potential $H_\mathrm{el}$, the inter-dot tunneling $H_t$ and the Zeeman effect $H_{z}^j$ in the left ($j=l$), center ($j=c$) and right ($j=r$) QD. In all three dots only the lowest orbital is considered. Explicitly,
\begin{equation}
H_\mathrm{el} = \sum_{j= c,l,r} \left( E_j n_j + \frac{U_{2j}}{2}\, n_j (n_j -1) \right) + U_1 (n_l n_c + n_c n_r),
\end{equation}
where $n_j=\sum_{\sigma}c_{j\sigma}^\dagger c_{j\sigma}$ denotes the total occupation number operator in QD $j$ and $c_{j\sigma}^{(\dagger)}$ annihilates (creates) an electron with spin $\sigma$ in QD $j$. The potentials $E_j$ can be tuned electrically. The Coulomb repulsion between electrons in adjacent dots, $U_1$, and in the same dot, $U_{2j}$, are determined by the inter-dot distance and the QD radius.

Tunneling between QDs is included with
\begin{equation}
H_t = \sum_{\sigma} t_{lc} \left( c_{l\sigma}^\dagger c_{c\sigma} + \mathrm{h.c.} \right) + t_{cr} \left( c_{c\sigma}^\dagger c_{r\sigma} + \mathrm{h.c.} \right).
\end{equation}
Here, $t_{lc(cr)}$ is the tunneling matrix element between the left and center (center and right) QD and h.c. denotes the Hermitian conjugate. Note that $H_t$ includes only spin conserving tunneling. Spin-orbit interaction (SOI) can lead to spin-flip tunneling \cite{RevModPhys.79.1217,doi:10.1146/annurev-conmatphys-030212-184248}, this is commented on in Sec.~\ref{sec_trans_grad}.

In QD $j$, the spin Hamiltonian is of the form $H_z^j= \vec{B}_{j} \cdot \vec{S}_j$, where $\vec{S}_j$ is the spin operator at site $j$. The local magnetic fields $\vec{B}_{j}$ are given in energy units \cite{RevModPhys.79.1217} and comprise a homogeneous external field $B_\mathrm{ext}\hat{\vec{z}}$ along the $z$-axis and potentially an inhomogeneous contribution from a static Overhauser field or a micromagnet \cite{RevModPhys.79.1217,EnrichedSilicon}. To quantify the inhomogeneity we define the longitudinal ($b_{zj}$) and transverse ($b_{xj}$) magnetic field differences for $j=l,c$, $\alpha=x,z$,
\begin{equation}
b_{\alpha j} = (\vec{B}_j - \vec{B}_r)_\alpha,\label{eq_def_gradients}
\end{equation}

For the remainder of this work we assume that $E_j$, $j=l,c,r$ are adjusted such that two electrons are confined to the TQD. Furthermore, we define $\eps=E_c - E_r$ and $\delta = E_l - E_c - U_1$ and assume that the TQD is operated in the regime $|\delta| \ll |\eps| + U_1 \ll U_{2c}, U_{2r}$
. Using the notation $(n_l,n_c,n_r)$ for the charge configuration, the operating regime allows two stable charge configurations, $(1,0,1)$ and $(0,1,1)$, i.e. the right QD is always occupied by one electron while the second electron can be either in the left or center QD. States with doubly occupied QDs, $(0,2,0)$ and $(0,0,2)$, are split off by a large spectral gap of the order $U_{2c(r)}-U_1$ and have very low occupation probability. Consequently, the Hamiltonian can be reduced to the low-energy subspace of states with single occupation. This can be accomplished by a Schrieffer-Wolff transformation \cite{SW_orig,BRAVYI20112793} and results in \cite{PhysRevB.97.085421}
\begin{eqnarray}
H'_\mathrm{TQD} &=&  \left[J \left( \boldsymbol{S}_c \cdot \boldsymbol{S}_r -\frac{1}{4} \right) +\boldsymbol{B}_c \cdot \boldsymbol{S}_c \right]\frac{1\!\! 1 - \tau_z}{2} \label{eq_Hexchange}\\
&& + \left(\delta + \boldsymbol{B}_l \cdot \boldsymbol{S}_l\right)\frac{1\!\! 1 + \tau_z}{2} + t_{lc} \tau_x + \boldsymbol{B}_r \cdot \boldsymbol{S}_r,\nonumber
\end{eqnarray}
where $\tau_z = |1,0,1\rangle \langle 1,0,1| - |0,1,1\rangle \langle 0,1,1|$ is the Pauli $z$ operator and $\tau_x = |1,0,1\rangle \langle 0,1,1| + \mathrm{h.c.}$ is the Pauli $x$ operator for the two available charge configurations. In the low energy subspace $U_1$ is but an offset of the left dot potential $\delta$ and the two-electron dynamics is captured in the exchange energy
\begin{equation}
J = \frac{2 t_{cr}^2 (U_{2c} + U_{2r} - 2U_1)}{(U_{2c} - U_1 + \eps)(U_{2r} - U_1 - \eps)}.\label{eq_defJ}
\end{equation}

The energy level diagram of $H'_\mathrm{TQD}$ near the $(1,0,1)\leftrightarrow (0,1,1)$ charge transition is depicted in Fig.~\ref{fig_systemsketch}(c).

The charge states $(1,1,0)$ and $(2,0,0)$ are neglected entirely in the derivation of $H'_\mathrm{TQD}$. This is justified only if $U_1 \gg \eps,\delta$, otherwise a small correction arises. Virtual tunneling into these states gives rise to a shift $\delta \to \delta_s = \delta + 2t_{cr}^2/(U_1+\eps+\delta)$. Furthermore a superexchange coupling $J_s$ between the electrons in the outer dots which is in leading order $\propto t_{lc}^2t_{cr}^2/U_C U_1^2$ which is typically much smaller than $J$.

The resonator is modeled as a single mode harmonic oscillator, $H_\mathrm{res}=\omega_0 a^\dagger a$ with annihilation (creation) operator $a^{(\dagger)}$, choosing $\hbar=1$. The electric field $\vec{E}$ couples to the dipole $e\vec{r}$ of the DQD via $H_\mathrm{dip}= e \vec{E}\cdot\vec{r}$ \cite{Cohen-Tannoudji,ScullyZubairy} where $e$ is the electron charge. In the present case this can also be written as $H_\mathrm{dip} = 2g_0(a+a^\dagger)\tau_z$ with $g_0 = e E_0 r_0$. The matrix elements include the projection $E_0$ of the electric field $\vec{E}$ to the DQD-axis and the distance $r_0$ between the left and center QD. The interaction must also be transformed into the low energy subspace, but if $|\eps| + U_1 \ll U_{2c}, U_{2r}$ then $H'_\mathrm{dip} \approx H_\mathrm{dip}$.

A comprehensive sketch of the system is shown in Fig.~\ref{fig_systemsketch}. We consider a setup where, additionally, a coherent driving field
\begin{equation}
H_p= i\sqrt{\kappa_1}\left( \ain e^{-i\omega_p t}a^\dagger - \ain^* e^{i\omega_p t}a  \right)
\end{equation}
with frequency $\omega_p$ and amplitude $|\ain |$ enters the cavity through port 1 (see Fig.~\ref{fig_systemsketch}b). 
At port 2 
the transmitted field $\aout$ is measured. Port $i=1,2$ has the leakage rate $\kappa_i$, the total cavity leakage rate is $\kappa=\kappa_1 + \kappa_2$. The total Hamiltonian is then
\begin{equation}
H = H'_\mathrm{TQD}+H_\mathrm{res}+H'_\mathrm{dip}+H_p
\end{equation}

Input-output theory \cite{PhysRevA.30.1386} is used to compute the stationary state of the output field $\aout = \sqrt{\kappa_2} a$ and the normalized transmission $T=|\aout/\ain|^2$ of  the resonator, following the lines of Refs.~\cite{PhysRevB.94.195305,PhysRevA.98.023849}. The Hamiltonian is transformed into the eigenbasis of $H'_\mathrm{TQD}$, defined by $U_\mathrm{TQD}H'_\mathrm{TQD}U_\mathrm{TQD}^\dagger=\mathrm{diag}\left(E_1,E_2,...\right)$, $E_n\leq E_{n+1}$, and further into a rotating frame to remove the time dependence from $H_p$. We apply a rotating wave approximation (RWA). We choose a rotating frame that allows to observe transitions between states adjacent in energy. Solving the quantum Langevin equations for $a$ and the TQD ladder operators yields \cite{PhysRevB.94.195305}
\begin{eqnarray}
\frac{\aout}{\ain} &=& \frac{-i \sqrt{\kappa_1 \kappa_2}}{\omega_0 - \omega_p - i \kappa / 2 + 2g_0 \sum_n d_{n,n+1} \chi_{n,n+1}},\label{eq_transmitivity}\\
\chi _{n,n+1} &=& \frac{-2g_0 d_{n+1,n}(p_n - p_{n+1})}{E_{n+1} - E_n - \omega_p - i \gamma /2},\label{eq_susceptibility}
\end{eqnarray}
where $\gamma$ is the dephasing rate of the DQD states and $d_{nm}$ are the matrix elements of $d=U_\mathrm{TQD}\tau_z U_\mathrm{TQD}^\dagger$. From Eq.~(\ref{eq_susceptibility}) it follows that the cavity transmission shows a dip if the TQD is tuned to an avoided crossing (AC) whose splitting matches $\omega_p$.

The TQD is assumed to have the finite temperature $T_\mathrm{dot}$, with the thermal population $p_n =\exp\left(-E_n / k_B T_\mathrm{dot}\right)/\sum_j \exp\left(-E_j / k_B T_\mathrm{dot}\right)$ in the $n$th eigenstate of $H_\mathrm{DQD}$. Here, $k_B$ is the Boltzmann constant.

\section{Proposed transmission-based measurement of the exchange\label{sec_simplest_case}}

It is well known that some Hamiltonian parameters that govern the single-electron dynamics in a quantum dot system can be extracted from the cavity transmission $T$ \cite{PhysRevB.94.195305,PhysRevLett.119.176803,Russ_2020,PhysRevB.98.161404,Hao2014,Borjans2021,Chen2021}. It is desirable to have a similarly simple way to characterize the exchange $J$ between two electrons in adjacent QDs in the $(0,1,1)$ charge configuration. As discussed in App.~\ref{app_11_scheme} the transmission in the $(0,1,1)$ regime with $t_{lc}=0$ carries information about the exchange $J$ that can be classically measured. However, there the cavity response has a visibility of $\lesssim 10^{-5}$ under realistic conditions since the dipole moment of the electron charge is very small in this regime, due to the small contribution of the $(0,2,0)$ and $(0,0,2)$ charge states.

To evade the problem of the small dipole moment we propose to sweep the electrostatic potential $\delta$ of the left dot through the $(1,0,1)\leftrightarrow (0,1,1)$ charge transition. The dipole moment of this transition allows for a sufficiently visible cavity response which depends on the two-electron spin state. This allows to extract the energy splitting of $J$ between the $(0,1,1)$ singlet and unpolarized triplet states, $|S^{cr}\rangle$ and $|T_0^{cr}\rangle$.

\begin{figure}
\includegraphics{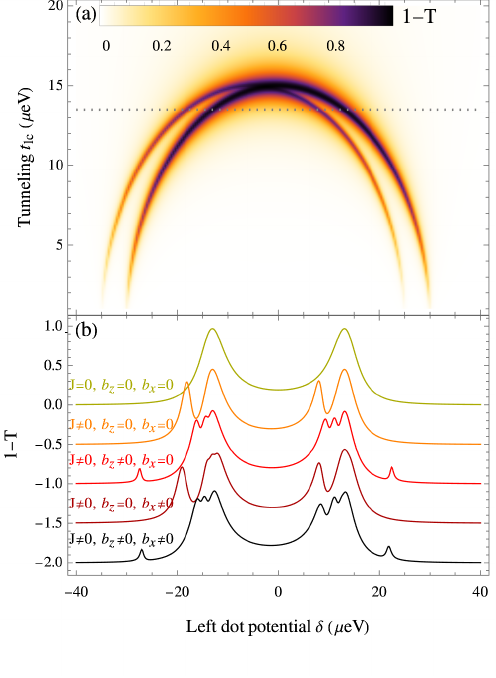}
\caption{Basic principle of the exchange measurement scheme.
(a) During sweeps of $\delta$ and $t_{lc}$ the avoided crossings of singlet and triplet states give rise to absorption probabilities $1-T$ which are split by the exchange $J$. The plot shows the example of vanishing gradients, $J=\SI{5}{\micro\electronvolt}$, $B_\mathrm{ext}=\SI{30}{\micro\electronvolt}$, $\eps=\SI{100}{\micro\electronvolt}$, $T_\mathrm{dot} = \SI{0.75}{\kelvin}$, $\omega_p = \omega_0 =\SI{30}{\micro\electronvolt}$, $g_0=\SI{0.2}{\micro\electronvolt}$, and $\gamma=\SI{0.5}{\micro\electronvolt}$, $\kappa_1 = \kappa_2 =\SI{0.0128}{\micro\electronvolt}$.
(b) Linecuts through panel (a) at $t_{lc} = \SI{13.5}{\micro\electronvolt}$ (dotted line) illustrate the effects of $J$ and magnetic gradients $b_{zl}=2b_{zc}=\SI{4}{\micro\electronvolt}$ and $b_{xl}=2b_{xc}=\SI{5}{\micro\electronvolt}$. For clarity, the curves are displaced by 0.5 each. The splitting of the $T_\pm$ responses due to $b_{xl(c)}$ cannot be resolved with this setting (dark red and black curves). The three lowest curves are obtained from numerical diagonalization of $H'_\mathrm{TQD}$.\label{fig_ex_novalley}}
\end{figure}

We first discuss the case without magnetic gradients, $b_{z,l}=b_{z,c}=0$, $b_{x,l}=b_{x,c}=0$. Here, it is straightforward to derive $T$ explicitly from Eqs.~(\ref{eq_transmitivity}) and (\ref{eq_susceptibility}), as shown in App.~\ref{app_T}. The cavity response has two contributions, one due to the tunneling between the singlet states at the two sites, the other due to the tunneling between the triplet states. As a function of $\delta$ and $t_{lc}$ the responses exhibit the characteristic arc shape of an AC (Fig.~\ref{fig_ex_novalley} (a)). During a sweep of the left dot potential $\delta$  two pairs of resonances are thus observed at
\begin{eqnarray}
\delta_{1\pm} &=& -J \pm \sqrt{\omega _p^2-4 t_{lc}^2},\label{eq_singletresponse}\\
\delta_{2\pm} &=& \pm \sqrt{\omega _p^2-4 t_{lc}^2}\label{eq_tripletresponse}.
\end{eqnarray}
Due to the finite exchange $J$ the transition between the singlets is shifted, directly revealing the value of $J$ [Eq.~(\ref{eq_singletresponse})]. This is illustrated by the gold ($J=0$) and orange ($J\neq 0$) curves in Fig.~\ref{fig_ex_novalley}(b).

In this simple case we envision an experiment where $1-T$ is measured as a function of $\delta$. From this profile the peaks $\delta_1$ and $\delta_2$ can be inferred and taking the distance between adjacent peaks $J = \delta_{2\pm}-\delta_{1\pm}$ can be directly read from the data independent from $t_{lc}$ and an offset of $\eps$. Alternatively, the distance between the outermost peaks can be inferred, $\delta_{2+}-\delta_{1-}$. This difference is not independent from $t_{lc}$ and $\omega_p$, thus a single-parameter fit of a measurement series as a function of $t_{lc}$ can be useful to minimize the uncertainty.

Note that the two arcs intersect at $\left(\delta, t_{lc}\right)=\left(-J/2,\frac{1}{4} \sqrt{4 \omega _p^2 - J^2}\right)$. If $2t_{lc}\approx \omega_p \gg J$ is chosen the transmission dips are hard to resolve individually. Thus, we recommend to choose $2t_{lc} < \omega_p$ for the measurement. Without the gradients no spin-flip processes are present and thus $B_\mathrm{ext}$ enters only via the populations. We recommend to chose $k_B T_\mathrm{dot} \gtrsim B_\mathrm{ext}$ for a significant population in the singlet state.

The result of this section is based on the Hamiltonian Eq.~(\ref{eq_Hexchange}) which relies on the assumption that the electrons cannot interact unless they occupy adjacent QDs. Realistically, it is possible that a superexchange $J_s$ couples electrons occupying the left and right dots, e.g. due to virtual tunneling to the $(1,1,0)$ and $(2,0,0)$ charge states. Repeating the previous derivation with such a small $J_s$ we find that in this case the difference between the exchange coupling is measured: The transition between the singlets is visible at $\delta_{1s\pm} \approx -(J-\frac{J_s}{4}) \pm \sqrt{\omega _p^2-4 t_{lc}^2}$ and the response associated with the triplet states remains at $\delta_{2\pm} \approx \pm \sqrt{\omega _p^2-4 t_{lc}^2}$.

\section{Discussion of magnetic gradients\label{sec_grad}}

In quantum information applications it may be required to include a micromagnet into the QD device device to perform fast gate operations \cite{PhysRevLett.96.047202,PhysRevB.83.121403,PhysRevB.97.085421,Zajac439,Landig2018} or to realize spin-photon coupling \cite{PhysRevB.86.035314,StrongCouplingPrinceton,Samkharadze1123}. 
A longitudinal gradient $b_{zl},b_{zc}$ mixes the singlet and unpolarized triplet states $T_0$ while a transverse gradient $b_{xl},b_{xc}$ allows spin-flip transitions to the spin-polarized triplets $T_\pm$.

\subsection{Longitudinal magnetic gradient\label{sec_long_grad}}

To discuss the role of the longitudinal gradient we assume $b_{xl}=b_{xc}=0$ and treat $b_{zl}/J,b_{zc}/J$ as a perturbation of $H'_\mathrm{TQD}$, Eq.~(\ref{eq_Hexchange}). We apply non-degenerate perturbation theory to derive the corrections up to second order. Two prime effects of $b_{zl},b_{zc}$ are found.

Due to the refined energy splitting between the $S$ and $T_0$ states the response from the tunneling of the singlets is shifted to
\begin{equation}
\delta_{1\pm}' \approx -J \pm \sqrt{\omega_p^2 - 4 t_{lc}^2} + \frac{2 b_{zl}^2}{J} - \frac{b_{zc}^2}{J} - \frac{2 b_{zl} b_{zc}}{J},
\end{equation}
the tunneling of the $T_0$ states is now observed at
\begin{equation}
\delta_{2\pm}' \approx \pm \sqrt{\omega_p^2 - 4 t_{lc}^2} - \frac{2 b_{zl}^2}{J} + \frac{b_{zc}^2}{J} + \frac{2 b_{zl} b_{zc}}{J}.
\end{equation}
The responses due to the tunneling of the $T_\pm$ states are not affected and still appear as specified by Eq.~(\ref{eq_tripletresponse}).

Furthermore, additional transmission dips can be observed, stemming from $S$-$T_0$ transitions. These appear near
\begin{eqnarray}
\delta_{3\pm}' &\approx & -\frac{J}{2} \pm \Bigg(\frac{2 b_{zl}^2}{J} - \frac{b_{zc}^2}{J} - \frac{2 b_{zl} b_{zc}}{J} \\
&& + \frac{|J+ 2 \omega_p|}{2} \sqrt{1+ \frac{(2 b_{zl} + b_{zc})^2 - (2t_{lc} - b_{zc})^2}{\omega_p (J + \omega_p)}}\Bigg),\nonumber \\
\delta_{4\pm}' &\approx & -\frac{J}{2} \pm \Bigg( \frac{2 b_{zl}^2}{J} - \frac{b_{zc}^2}{J} - \frac{2 b_{zl} b_{zc}}{J} \\
&& + \frac{|J - 2 \omega_p|}{2} \sqrt{1 + \frac{(b_{zl} + b_{zc})^2 + (2t_{lc} + b_{zc})^2}{\omega_p (J - \omega_p)}} \Bigg). \nonumber
\end{eqnarray}
Both effects are visible in the example of Fig.~\ref{fig_ex_novalley}(b) (light red curve).

The example Fig.~\ref{fig_ex_novalley}(b) also highlights that a longitudinal magnetic gradient can potentially be detrimental. If
\begin{equation}
\left|\frac{J}{2} - \frac{2 b_{zl}^2}{J} + \frac{b_{zc}^2}{J} + \frac{2 b_{zl} b_{zc}}{J}\right| \lesssim \gamma/2
\end{equation}
the responses from singlet and triplet states cannot be distinguished clearly.

We envision a reliable identification of the exchange coupling with longitudinal gradient by measuring $1-T$ as a function of $\delta$ for different values of $t_{lc}$. From each of these traces the position of the outermost peaks $\delta_{3\pm}'$ can be inferred. Provided that $b_{zl}$, $b_{zc}$ are known, a single-parameter fit of $\delta_{3+}'-\delta_{3-}'$ to the distance between the two peaks as a function of $t_{lc}$ reveals $J$. The limitation to this protocol is the requirement to clearly distinguish the outermost peaks from the others and it fails if $|\delta_{3-}'-\delta_{1-}'|\lesssim \gamma/2$. Note that in this case a superexchange $J_s$ will lead to a small correction of $\delta_{2\pm}'$. To first order the correction will be $\propto J_s$.

The longitudinal gradient can be measured beforehand by measuring the Zeeman splitting in each dot and taking their differences, according to Eq.(3). Alternatively, a multi-parameter fit of $\delta_{3+}'-\delta_{3-}'$ to the data can can be used to simultaneously estimate $J$ and the gradient, enough data points provided.

\subsection{Transverse magnetic gradient\label{sec_trans_grad}}

Similarly to Sec.~\ref{sec_long_grad}, here, we assume that $b_{zl}=b_{zc} =0$ and treat $b_{xl}, b_{xc}$ with second order non-degenerate perturbation theory. The resulting expressions are expanded around the charge transition, assuming $\left(b_{xc}-b_{xl}\right)^2 \ll \omega _p^2 - 4 t_{lc}^2$.

Due to the admixture of $T_\pm$ states the responses due to the tunneling of the singlets is shifted as a function of $B_\mathrm{ext}$ to
\begin{eqnarray}
\delta_{1\pm}'' &\approx & - J \pm \Bigg[\left(\omega_p + \frac{4 t_{lc} (b_{xc} - b_{xl})^2 (4 B_\mathrm{ext}^2 + J^2 - 16 t_{lc}^2)}{(4 B_\mathrm{ext}^2 + J^2 - 16 t_{lc}^2)^2 - (4 B_\mathrm{ext} J)^2} \right)^2 \nonumber \\
&& - 4 t_{lc}^2 \Bigg]^{1/2}\label{eq_bx_correctionS}
\end{eqnarray}
while the tunneling of the $T_\pm$ states is now observed in separate transmission dips at
\begin{equation}
\delta_{2\pm}'' \approx \pm \sqrt{\left( \omega_p + \frac{2 (b_{xc} - b_{bz})^2 t_{lc}}{(J \pm 2B_\mathrm{ext})^2 - (4 t_{lc})^2} \right)^2 - 4t_{lc}^2}.\label{eq_bx_correctionT}
\end{equation}
The response due to the tunneling of the $T_0$ triplet is not affected in this case and remains as given by Eq.~(\ref{eq_tripletresponse}).
Note that the corrections in Eqs.~(\ref{eq_bx_correctionS}) and (\ref{eq_bx_correctionT}) become singular for $B_\mathrm{ext} = \pm 2 t_{lc} \pm J/2$. There, the non-degenerate perturbation theory breaks down.

Furthermore, $b_{xl},b_{xc}$ allow for a number of additional spin-flip transitions. The associated dipole moments are $\propto (b_{xl}/2B_\mathrm{ext})^2$, however. Thus, we propose to choose a large magnetic field $B _\mathrm{ext} \gg 2t_{lc}$ for the measurement. This eliminates undesired responses and makes sure the analytical results from perturbation theory can be applied.

The effects of the transverse magnetic gradient alone and in conjunction with the longitudinal gradient are shown in Fig.~\ref{fig_ex_novalley}(b) (dark red and black curves). In this example Eqs.~(\ref{eq_bx_correctionS}) and (\ref{eq_bx_correctionT}) are only a coarse approximation since $B_\mathrm{ext}$ is close to $2t_{lc}+J$.

If both gradients, longitudinal and transverse, are present, the same experimental procedure as with only the longitudinal gradient (Sec.~\ref{sec_long_grad}) can be applied since the correction to the additional features is negligible. In case of a transverse gradient alone the exchange $J$ can still be found by inferring the distance between the outermost peaks, $\delta_{1-}''$ and $\delta_{2-}''$, and a single-parameter fit to our equations if $b_{xl}$ and $b_{xc}$ are known. This can result in an enhanced uncertainty of $J$, however, if the peaks from the two polarized triplets are merged into one broadened peak as in Fig.~\ref{fig_ex_novalley}(b), i.e. if their separation is comparable to or smaller than $\gamma/2$.

The required knowledge about $b_{xl}$ and $b_{xc}$ can be estimated in a preceding experiment where the spin-flip tunneling between the QDs is measured in the weak coupling regime with a single electron \cite{PhysRevB.102.195418}. The case with $b_{xl}$ and $b_{xc}$ alone is expected to be irrelevant for realistic applications with a micromagnet, however.

Another physical process that can introduce spin-flip processes is spin-orbit interaction (SOI) \cite{PhysRevLett.110.196803,PhysRevB.85.125312,NatCommun2.556,doi:10.1146/annurev-conmatphys-030212-184248}. To include these processes we use a modified Hamiltonian $H_\mathrm{TQD} \to H_\mathrm{TQD}^\mathrm{SOI}$ (Eq.~\ref{eq_Hexchange}) with a complex spin-flip tunneling terms $f_{ij}$,
\begin{eqnarray}
H_t \to H_t^\mathrm{SOI} &=& \sum_{\sigma,\sigma'}  \Big[ \delta_{\sigma'\sigma} t_{lc} c_{l\sigma}^\dagger c_{c\sigma} \\
&+& (1-\delta_{\sigma'\sigma}) f_{lc} c_{l\sigma}^\dagger c_{c\sigma'} + \mathrm{h.c.} \Big] + [(\mathrm{lc})\to (\mathrm{cr})].\nonumber
\end{eqnarray}

By treating $f_{ij}$ as perturbation and including terms up to second order we find that the effect of the SOI is of similar form as the transverse magnetic gradient. The explicit expressions are presented in App.~\ref{app_SOI}. Note that in the presence of the spin-flip terms $J$ is not given by Eq.~(\ref{eq_defJ}). The SOI can furthermore contribute to the position-dependent part of the spin splitting \cite{PhysRevB.98.245424}. This effect can be directly incorporated into $b_{zl}, b_{zc}$.

\section{Valley degree-of-freedom\label{sec_valley}}

For spin qubits realized in silicon the valley pseudospin \cite{doi:10.1063/1.1637718,PhysRevLett.88.027903,PhysRevB.81.115324,Hollmann2019} can be described with a pseudospin operator $\vec{V}_j$ with ladder operators $V_{j\pm}=V_{j x}\pm i V_{j y}$. In each singly occupied dot $j$ the valley Hamiltonian is given by $H_v^j = \Delta_j e^{i \varphi_j} V_{j+} + \mathrm{h.c.}$ \cite{PhysRevB.75.115318,PhysRevB.82.155312}. The valley splitting $\Delta_j$ and phase $\varphi_j$ can differ between the dots \cite{PhysRevB.80.081305,PhysRevB.84.155320,PhysRevResearch.2.043180}.  In the valley eigenbasis of all dots the valley phase differences $\df_{lc}= (\varphi_l - \varphi_c)/2$ and $\df_{cr}= (\varphi_c - \varphi_r)/2$ can be viewed as the angles between the valley pseudospins in adjacent dots and parametrize the ratio of of valley conserving ($t_{ij} \cos\df_{ij}$) and valley-flip tunneling ($t_{ij} \sin\df_{ij}$) between these dots \cite{PhysRevB.94.195305,Russ_2020}.

The low-energy Hamiltonian is analogous to Eq.~(\ref{eq_Hexchange}).
\begin{eqnarray}
H_\mathrm{TQD}^{\prime v} &=& H_{(101)}^v \frac{1\!\! 1 + \tau_z}{2} + H_{(011)}^v \frac{1\!\! 1 - \tau_z}{2} + t_{lc} \tau_x, \\
H_{(101)}^v &=& \delta + \sum_{j=l,r} \left( \vec{B}_j \cdot \vec{S}_j + H_v^j \right).
\end{eqnarray}
The exchange contribution in $H_{(011)}^v$ strongly depends on the presence of various interaction terms and its structure depends on the assumptions made \cite{David_2018}. For our Hamiltonian with the interaction terms introduced in Sec.~\ref{sec_TQD_model} and assuming $|B_{c(r)}| \gg  b_{zc}$ and $|\Delta_{c(r)}| \gg |\Delta_c - \Delta_r |$ and $B_\mathrm{ext}, b_{zc}, b_{xc} \neq 0$, we find
\begin{eqnarray}
H_{(011)}^v &\approx & \sum_{j=c,r} \left( \vec{B}_j \cdot \vec{S}_j + H_v^j \right) + \frac{J}{8}\Big[(\vec{S}_c \cdot \vec{S}_r) (\vec{V}_c \cdot \vec{V}_r) \nonumber\\
&& + \vec{S}_c \cdot \vec{S}_r + \vec{V}_c \cdot \vec{V}_r + 8 \Big(\frac{1}{4} + \vec{S}_c \cdot \vec{S}_r \nonumber\\
&& - 2 S_{cy} S_{ry} \Big)\Big(\frac{1}{4} - \vec{V}_c \cdot \vec{V}_r \Big) - 3\Big].
\end{eqnarray}
The low-energy Hamiltonian has 32 relevant basis states, forming six supersinglets and ten supertriplets in each charge configuration \cite{PhysRevB.82.155424,Rohling_2012}.

First, we consider the effect of the lifted valley degeneracy for the limit of a large valley splitting which is comparable to the Zeeman splitting, $\Delta_j \approx B_\mathrm{ext} \gg J$ and $4 t_{lc} < \Delta_l + \Delta_c - \left|\Delta_l - \Delta _c\right|$. In this limit it is possible to treat $J$ and the magnetic gradients as perturbations and approximate the eigenenergies of $H_\mathrm{TQD}^{\prime v}$ near the ACs.

Knowledge about the valley phase differences and thus the occurrence of valley-flip tunneling is of vital importance for the interpretation of the results. The splitting of the ACs at the charge transition is determined by $\df_{lc}$. On the other hand, $\df_{cr}$ determines which $(0,1,1)$ states can couple to the $(0,2,0),(0,0,2)$ subspace and are thus shifted in energy by the exchange interaction.

We find that the valley-conserving tunneling between the left and center dot gives rise to up to twelve pairs of transmission dips with a dipole moment $\propto t_{lc} \cos(\delta\varphi_{lc})$. The tunneling between states without spin polarization is observed in the cavity response near
\begin{equation}
\delta_{1v} \approx \pm \frac{\Delta_c - \Delta_l}{2} -\frac{J}{2} \pm (b_{zc} - b_{zl}) \pm \sqrt{\omega_p^2 - 4 t_{lc}^2 \cos^2(\delta\varphi_{lc})}
\end{equation}
while the spin-polarized states are observed near
\begin{eqnarray}
\delta_{2v} &\approx & \pm \frac{\Delta_c - \Delta_l}{2} -\frac{J}{8}[1 \pm \cos(\delta\varphi_{cr})] \pm \frac{b_{xc}^2 - b_{xl}^2}{2B_\mathrm{ext}} \nonumber\\
&& \pm \sqrt{\omega_p^2 - 4 t_{lc}^2 \cos^2(\delta\varphi_{lc})}.
\end{eqnarray}

\begin{figure}
\includegraphics{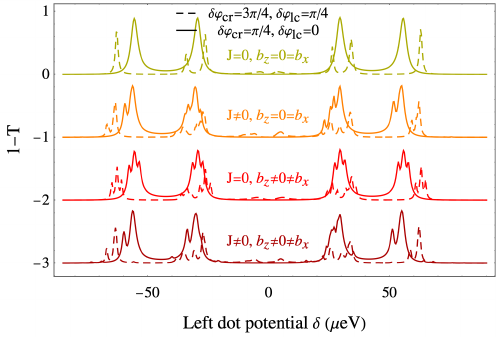}
\caption{Cavity absorption $1-T$ during a sweep of $\delta$ with lifted valley degeneracy of $\Delta_l = \SI{45}{\micro\electronvolt}$, $\Delta_c = \SI{40}{\micro\electronvolt}$, $\Delta_r = \SI{50}{\micro\electronvolt}$ and two different combinations of valley phase differences $\df_{lc}$, $\df_{cr}$ (dashed and solid curves). As can be seen, the proposed measurement scheme is reliable in the presence of a valley splitting as well and the effect of the gradients can be expected to be comparable to the case without valley. The precise transmission profile strongly depends on the different tunneling matrix elements parametrized by the valley phase differences. Here, $B_\mathrm{ext} = \SI{50}{\micro\electronvolt}$, while all other parameters are the same as in Fig.~\ref{fig_ex_novalley}(b). For clarity the curves are offset by 1 each.\label{fig_valley}}
\end{figure}

Analogously, the valley-flip tunneling between $l$ and $c$ gives rise to responses with dipole moment $\propto t_{lc} \sin(\delta\varphi_{lc})$ near
\begin{equation}
\delta_{3v} \approx \pm \frac{\Delta_c + \Delta_l}{2} -\frac{J}{2} \pm (b_{zc} - b_{zl}) \pm \sqrt{\omega_p^2 - 4 t_{lc}^2 \sin^2(\delta\varphi_{lc})}
\end{equation}
and also near
\begin{eqnarray}
\delta_{4v} &\approx & \pm \frac{\Delta_c + \Delta_l}{2} -\frac{J}{8}[1 \pm \cos(\delta\varphi_{cr})] \pm \frac{b_{xc}^2 - b_{xl}^2}{2B_\mathrm{ext}} \nonumber \\
&& \pm \sqrt{\omega_p^2 - 4 t_{lc}^2 \sin^2(\delta\varphi_{lc})}
\end{eqnarray}
These results are illustrated in Fig.~\ref{fig_valley}.

In the opposite limit of a small valley splitting, $\Delta_j \approx J \ll B_\mathrm{ext}$ we treat $\Delta_j$ and $J$ as perturbations. As a simplification, only the lowest spin state is considered, which is justified if $B_\mathrm{ext} \gg k_B T_\mathrm{dot}$.

Due to the valley-conserving tunneling between the left and center QD up to four arc-shaped transmission dips emerge in the $\delta$-$t_{lc}$ plane, when the condition
\begin{eqnarray}
\omega_p \sqrt{16 t_{lc}^2 + \delta^2} &\approx & 8t_{lc}^2 + \delta \Big\{ -\delta - \frac{J}{8}\left[ 1 \pm \cos(2\delta\varphi_{cr})\right] \nonumber\\
&& \pm \frac{1}{2} \left[\Delta_c - \Delta_c \cos(2 \delta\varphi_{lc})\right] \Big\}
\end{eqnarray}
is satisfied. The valley-flip tunneling similarly gives rise to cavity responses if
\begin{eqnarray}
\omega_p &\approx & \frac{16 t_{lc}^2 + \delta (\delta + J/4)}{\sqrt{16 t_{lc}^2 + \delta^2}} + \frac{J}{8} [1 \pm \cos(2\delta\varphi_{cr})] \nonumber \\
&& \pm \frac{1}{2} [\Delta_c + \Delta_l \cos(2 \delta\varphi_{lc})].
\end{eqnarray}
Consequently, the exchange interaction can be identified from both, a sweep of $\delta$ and $t_{lc}$.

The measurement procedure for the exchange $J$ can be completed similar to the simple case of Sec.~\ref{sec_simplest_case}, inferring distance between the outermost peaks along the $\delta$-axis as a function of $t_{lc}$. A single-parameter fit then reveals $J$, if all other parameters are known. However, due to the strong dependence of peak position and dipole moment on the valley phase differences there are no universal instructions which peaks those will be. This is best determined case-by-case by using the equations from this section, prior knowledge about the valley parameters and an estimated range of $J$.

Therefore, before $J$ can be measured, the valley Hamiltonian should be estimated. This can be accomplished with well-known techniques described in Refs. \cite{PhysRevB.94.195305,PhysRevLett.119.176803,Borjans2021} using the same microwave resonator and a single electron.

\section{Conclusions\label{sec_conclusion}}

In this article we have proposed a scheme to measure the exchange coupling $J$ between two electrons in neighboring QDs from the transmission of a dipole-coupled microwave resonator. The exchange interaction between adjacent QDs emerges in a regime where charge transitions are extremely unlikely, resulting in a very low dipole moment. Our proposal circumvents this hindrance by introducing an empty third QD. The required dipole moment is obtained by sweeping through a charge transition where one electron can tunnel into the additional (left) QD. The relative position of the observed transmission dips reveals the value of $J$.

Exact analytical expressions for the transmission $T$ and the position of the transmission dips during a sweep of the left dot potential $\delta$ were derived. Furthermore, we applied perturbation theory up to second order to discuss corrections due to magnetic field gradients and weak spin-orbit interaction. A transverse magnetic field gradient $b_{xj}$ has only small effects if the external magnetic field $B_\mathrm{ext}$ is sufficiently high. A longitudinal gradient  $b_{zj}$, however, can obstruct the measurement, since it alters the singlet-triplet splitting.

The proposed measurement scheme also works in the case of a lifted valley degeneracy, e.g. in silicon QDs. Approximate expressions for the position of the transmission dips are presented in the limits of large and small valley splitting. In both cases the valley phase differences $\df_{ij}$ have a crucial role. The phase differences parametrize the valley-conserving and valley-flip tunneling and thus determine which transitions couple to the cavity field.

Our results can be applied to simplify and speed up the characterization of the short-range interaction between spin qubits, of multi-spin qubit devices and the interaction in longer spin chains. With the addition of the estimation of $J$ to their range of applications microwave resonators become even more significant a component of spin qubit devices.


\begin{acknowledgments}
We thank Jonas Mielke, Amin Hosseinkhani and Jason R. Petta for helpful discussions. Florian Ginzel acknowledges a scholarship from the Stiftung der Deutschen Wirtschaft (sdw) which made this work possible. This work has been supported by the ARO grant number W911NF-15-1-0149.
\end{acknowledgments}


\appendix
\section{Exchange measurement scheme with only two dots\label{app_11_scheme}}

If the left QD is decoupled from the rest of the system ($t_{lc}=0$) and the remaining double quantum dot (DQD) is in the $(1,1)$ configuration, the low-energy Hamiltonian $H'_\mathrm{DQD} = H_z^c + H_z^r + J (\vec{S}_c \cdot \vec{S}_r +1/4)$ is readily diagonalized which can be used to compute $T$ from Eq.s~(\ref{eq_transmitivity}),(\ref{eq_susceptibility}). Choosing $\omega_p = \omega_0 \gg b_{xc},b_{zc} \neq 0$ and sweeping $B_\mathrm{ext}$, a transmission dip will be observed at $B_\mathrm{resp}$ in response to transitions between the two lowest energy eigenstates and it is
\begin{equation}
J \approx B_\mathrm{resp} - \omega_p + \frac{b_{zc}^2}{\omega_p - B_\mathrm{resp} }.\label{eq_J_DQD}
\end{equation}

However, the leading contribution to the dipole moment associated with this transition is of the order of $b_{xc} t_{cr}^2/(\min (U_{2r},U_{2c}) - U_1-|\eps|)^3$, resulting in an extremely low visibility of the corresponding cavity response. An analogous result can be obtained with lifted valley degeneracy $\Delta_j \gg J$, but the dipole moment and visibility are of the same order of magnitude.

\section{Explicit expression for $T$\label{app_T}}

Without magnetic gradients the transmission according to equation Eq.~(\ref{eq_transmitivity}) can be directly computed,
\begin{equation}
T=\left|\frac{a_{\mathrm{out}}}{a_{\mathrm{in}}}\right|^2=\left|\frac{- i \sqrt{\kappa _1 \kappa _2}}{\left(\omega _0-\omega _r\right) -\frac{i \kappa }{2}+2 g_0 \left[d_S(J)\chi _S + d_T \chi _T\right]}\right|^2.
\end{equation}
The two contributions to the cavity response,
\begin{eqnarray}
\chi_S &=& \frac{2g_0 d_S(J)}{a_J/2 - \omega_p - i \gamma/2}\; P_S, \\
\chi_T &=& \frac{2g_0 d_T}{a/2 - \omega_p - i \gamma/2} \, P_T
\end{eqnarray}
are stemming from the singlet (S) and triplet (T) states. We have defined $a_J = \sqrt{(4t_{lc})^2 + (2 J + 2 \delta)^2}$ and $a = \sqrt{(4t_{lc})^2 + (2 \delta)^2}$. The associated dipole moments are
\begin{eqnarray}
d_S (J) &=& - \prod_{\mu =\pm 1} \frac{\mu a_J/2 - J - \delta}{\sqrt{4t_{lc}^2 + (\mu a_J/2 - J - \delta)^2}}, \\
 d_T &=& - \prod_{\mu=\pm 1} \frac{\mu a/2 - \delta}{\sqrt{4t_{lc}^2 + (\mu a/2 - \delta)^2}}
\end{eqnarray}
and the populations in thermal equilibrium are
\begin{widetext}
\begin{eqnarray}
P_S &=& \frac{\sum_{\mu = \pm 1} \mu e^{-(\mu a_J/2 - J + \delta)/2 k_B T_\mathrm{dot}}}
{\sum_{\mu = \pm 1} \Big( e^{-(\mu a_J/2 - J + \delta)/2 k_B T_\mathrm{dot}} 
+ \sum_{\nu = -1}^1 e^{-(\mu a/2 + \nu B_\mathrm{ext} + \delta)/2 k_B T_\mathrm{dot}} \Big)},\\
P_T &=& \frac{\sum_{\mu = \pm 1} \sum_{\nu = -1}^1 \mu e^{-(\mu a/2 + \nu B_\mathrm{ext} + \delta)/2 k_B T_\mathrm{dot}} \Big)}
{\sum_{\mu = \pm 1} \Big( e^{-(\mu a_J/2 - J + \delta)/2 k_B T_\mathrm{dot}} 
+ \sum_{\nu = -1}^1 e^{-(\mu a/2 + \nu B_\mathrm{ext} + \delta)/2 k_B T_\mathrm{dot}} \Big)}.
\end{eqnarray}

\section{Corrections from SOI\label{app_SOI}}

The complex tunneling and spin-flip terms due to the SOI shift the response associated with the singlets to
\begin{equation}
\delta_1''' \approx - J t_{cr}^2/n \pm \sqrt{\left( \omega_p - \frac{|f_{lc}|}{4} \sum_{\mu,\nu = \pm 1} \frac{1}{2 \mu B_\mathrm{ext} + 4 t_{lc} + \nu J t_{cr}^2/n} \right)^2 - 4 t_{lc}^2},
\end{equation}
where $n=t_{cr}^2+|f_{cr}|^2$. The cavity responses associated with the $T_\pm$ states are shifted to
\begin{equation}
\delta_{2a}''' \approx \pm \sqrt{\left[ \omega_p - \frac{t_{lc} |f_{lc}|^2}{2} \left( \frac{1}{B_\mathrm{ext}^2 - 4 t_{lc}^2} + \frac{1}{(\pm B_\mathrm{ext} + J t_{cr}^2/2n)^2 - 4 t_{lc}^2} \right)\right]^2 - 4 t_{lc}^2}
\end{equation}
\end{widetext}
Unlike the transverse magnetic gradient, SOI introduces a small correction to the position of the cavity response associated with the $T_0$ states,
\begin{equation}
\delta_{2b}''' \approx \pm \sqrt{\left( \omega_p + \frac{t_{lc} |f_{lc}|^2}{B_\mathrm{ext}^2 - 4 t_{lc}^2} \right)^2 - 4 t_{lc}^2}
\end{equation}

Cavity responses due to additional transitions allowed for by the SOI can be neglected for large magnetic field similar to the case of the transverse magnetic gradient.

\bibliography{cavity-exchange_lit.bib}

\end{document}